\documentclass[twocolumn]{jpsj3}
\usepackage{txfonts}
\usepackage{mathtools}
\usepackage {graphicx}
\usepackage {multirow}
\usepackage{color}

\title{Nuclear Spin Relaxation Time due to the Orbital Currents in Dirac Electron Systems}

\author{Tomoki Hirosawa, Hideaki Maebashi, and Masao Ogata}
\inst{Department of Physics, University of Tokyo, Bunkyo, Tokyo 113-0033, Japan 
} 

\abst{The nuclear spin relaxation time $T_1$ is calculated taking account of the contributions from orbital currents of Dirac electrons. We consider a simple model of non-interacting Dirac electron gas in the three-dimensional bulk system. The obtained result shows $T^3$ dependence of $1/T_1$ at temperatures $T$ above the energy gap. This temperature dependence agrees qualitatively with the recent $\beta$-NMR experiment on the bulk of the topological insulator $\mathrm{Bi}_{0.9}\mathrm{Sb}_{0.1}$. }

\begin{document}

\maketitle

In condensed matter systems, the speed of electrons is much slower than the speed of light. However, it is known that a Hamiltonian of narrow-gap materials such as Bi can be written in a form similar to the Dirac Hamiltonian with a renormalized velocity in the solid. \cite{wolff_matrix_1964}Recently,  the polarization tensor of Dirac electrons has been reformulated in the context of condensed matter systems.\cite{Maebashi,weldon,bechler}  In the present paper, we apply this polarization tensor to calculate the nuclear spin relaxation time.

Firstly, we would like to note the intriguing property of the static susceptibility of Dirac electron systems, which was experimentally found and explained in terms of the orbital magnetism. The susceptibility of bismuth was shown to have the largest peak when the chemical potential is located in the band gap.\cite{Wehrli} This is not consistent with the Laudau-Peierls orbital susceptibility, and we need the interband contribution to understand this phenomenon. \cite{fukuyama_kubo,fukuyama_theory_1971,fukuyama_anomalous_2007,nakamura_orbital_2007} This has led us to recognize the importance of orbital magnetism in the Dirac electron systems.\cite{mizoguchi} 

The nuclear spin relaxation time $T_1$ is determined by the interaction between the nucleus and electrons. As the relaxation of nuclear spin occurs due to the fluctuation of effective magnetic field around the core, the relaxation time under the external magnetic field $H_0$ in the $z$-direction can be written in general\cite{chklovskii_relaxation_1992}:
\begin{equation}
\frac{1}{T_1}=\lim_{\omega\to0}\frac{\gamma^2_n\, \mu_0^2}{2}\times\frac{k_B T}{\hbar \omega}\mathrm{Im}\left[\mathopen|<\delta\mathbf{H}_x(\mathbf{r},\omega)>\mathclose|^2+\mathopen|<\delta\mathbf{H}_y(\mathbf{r},\omega)>\mathclose|^2\right],
\label{relaxation}
\end{equation}
where $\delta\mathbf{H}_{x,\, y}(\mathbf{r},\omega)$ denotes the effective magnetic field of electrons on the nuclear spin including contributions from the electron spin and the orbital current, $\mu_0$ is the vacuum permeability, and $\gamma_n$ is the gyromagnetic ratio for a nucleus.

As the rigorous evaluation of the above formula is very complicated, it is common to consider only a contribution from the Fermi contact interaction $H_{FC}=\frac{2}{3}\mu_0\gamma_n\gamma_e\hbar^2\mathbf{I}\cdot\mathbf{S}\delta(\mathbf{r})$. For metals and semimetals, it generally gives a reasonable prediction of experimental results. The resulting linear dependence of $1/T_1$ on the temperature is known as the Korringa relation\cite{korringa_nuclear_1950}:
\[T_1T K_s^2=\frac{\hbar}{4\pi k_B}\left(\frac{\gamma_e}{\gamma_n}\right)^2,\] where $K_s$ denotes the Knight shift. 
Although the modification of the numerical prefactor is required to account for the long-range interaction between electrons\cite{moriya_effect_1963}, it has successfully explained a wide range of NMR experiments on metals. For Dirac electron systems, however, it is known to show the large orbital diamagnetism in comparison to the spin susceptibility.\cite{fuseya_transport_2015} Thus, the orbital contribution to 1/$T_1$ could also be significant. Here, the orbital contribution means the contribution to $1/T_1$ due to the fluctuation of the magnetic field on nuclei induced by orbital currents of electrons, which is distinct from the Fermi contact interaction. In this paper, we investigate this contribution. 

The NMR studies on Dirac electron systems were conducted for several materials. Interestingly, the temperature dependence of 1/$T_1$ does not seem to follow the Korringa relation.
One example is a gapless quasi-two-dimensional Dirac electron system $\alpha$-(BEDT-TTF)$_2 \mathrm{I}_3$. The measured value of 1/$T_1$ showed $T^3$ dependence, which was explained by two-dimensional electronic density of states.\cite{katayama,miyagawa, hirata} Another example is a topological insulator $\mathrm{Bi}_{0.9}\mathrm{Sb}_{0.1}$, which is a three-dimensional gapped Dirac electron system in the bulk.\cite{hsieh_topological_2008} The $\beta$-NMR measurement showed a nonlinear increase in $1/T_1$ of $\mathrm{Bi}_{0.9}\mathrm{Sb}_{0.1}$, observed sharply at 225K. \cite{macfarlane_beta-detected_2014} In the following, we discuss that this result could be understood by orbital contributions of Dirac electrons. 
More recently, $T^3$ dependence was reported in NQR measurement on a Weyl semimetal TaP.\cite{NQR_TaP} The result for TaP was explained by the theoretical calculation of the orbital effect of the Weyl Hamiltonian.\cite{weyl} In the present paper, we take a different approach to obtain the rigorous expression for the Dirac Hamiltonian.

Our formulation follows the argument by Chklovskii and Lee.\cite{chklovskii_relaxation_1992} The nuclear relaxation time is determined by the fluctuation of the magnetic field, which can be related to the orbital current of electrons by Maxwell's equations in the momentum space. 
\begin{equation}
H_x(\mathbf{q},\omega)=i\frac{q_y j_z(\mathbf{q},\omega)-q_z j_y(\mathbf{q},\omega)}{q^2}.
\label{field}
\end{equation}
The expression for $H_y$ is given by its cyclic permutation. The momentum representation of (\ref{relaxation}) becomes
\begin{equation}
\begin{multlined}
\frac{1}{T_1}=\lim_{\omega\to0}\frac{\gamma^2_n \mu_0^2}{2} \frac{k_B T}{\hbar\omega}\int_{-\infty}^{\infty}\frac{d\mathbf{q}d\mathbf{q'}}{(2\pi)^6}\mathrm{Im}[\mathopen|<\delta\mathbf{H}_x(\mathbf{q},\omega)\delta\mathbf{H}_x(\mathbf{q'},\omega)>\mathclose| \\
+\mathopen|<\delta\mathbf{H}_y(\mathbf{q},\omega)\delta\mathbf{H}_y(\mathbf{q'},\omega)>\mathclose|].
\label{fourier}
\end{multlined}
\end{equation}
Assuming the translational invariance for a bulk system, we obtain
\begin{equation}
\begin{multlined}
<\delta\mathbf{H}_x(\mathbf{q},\omega)\delta\mathbf{H}_x(\mathbf{-q},\omega)>=\\
-\frac{1}{q^4}[-q_z^2 \mathopen<j_y(\mathbf{q},\omega) j_y (-\mathbf{q},\omega)\mathclose>-q_y^2\mathopen<j_z(\mathbf{q},\omega)j_z(-\mathbf{q},\omega)\mathclose> \\
+q_z q_y\mathopen<j_y(\mathbf{q},\omega)j_z(-\mathbf{q},\omega)\mathclose>+q_y q_z\mathopen<j_z(\mathbf{q},\omega)j_y(-\mathbf{q},\omega)\mathclose>],
\label{Hx}
\end{multlined}
\end{equation}
\begin{equation}
\begin{multlined}
<\delta\mathbf{H}_y(\mathbf{q},\omega)\delta\mathbf{H}_y(\mathbf{-q},\omega)>=\\
-\frac{1}{q^4}[-q_x^2 \mathopen<j_z(\mathbf{q},\omega) j_z (-\mathbf{q},\omega)\mathclose>-q_z^2\mathopen<j_x(\mathbf{q},\omega)j_x(-\mathbf{q},\omega)\mathclose> \\
+q_x q_z\mathopen<j_z(\mathbf{q},\omega)j_x(-\mathbf{q},\omega)\mathclose>+q_z q_x\mathopen<j_x(\mathbf{q},\omega)j_z(-\mathbf{q},\omega)\mathclose>].
\label{Hy}
\end{multlined}
\end{equation}

In the case of purely two-dimensional systems, $j_z(\mathbf{q},\omega)=q_z=0$ and all terms in Eqs. (\ref{Hx}) and (\ref{Hy}) vanish. For the quasi-two-dimensional system, the current flow is mostly confined in a two-dimensional plane, i.e., $j_z(\mathbf{q},\omega)=0$, but it has a periodic structure in the $z$-direction. As $q_z$ is well-defined in this system, terms such as $q_z^2 \mathopen<j_y(\mathbf{q},\omega) j_y (-\mathbf{q},\omega)\mathclose>$ remains finite. Hence, the orbital contribution is only finite for three-dimensional and quasi-two-dimensional systems. In the present work, we discuss the ordinary three-dimensional system. 

As introduced by Wolff, we could obtain the effective Hamiltonian for Bismuth by taking account of two bands near the small band gap. The isotropic three-dimensional Wolff Hamiltonian can be written as the Dirac Hamiltonian. \cite{fuseya_transport_2015}
\[\mathbf{H}=
\left(
\begin{array}{rr}
mc^{*2} & c^* \mathbf{k}\cdot \mathbf{\sigma} \\
c^* \mathbf{k}\cdot \mathbf{\sigma} & -mc^{*2}
\end{array}
\right),
\]
where $c^*$ denotes the velocity of Dirac electrons in the solid and $\sigma$ is the $2\times 2$ Pauli matrices. Thus, $\mathbf{H}$ is a $4\times 4$ Hamiltonian. For this Hamiltonian, the polarization tensor can be obtained by calculating the current-current correlation function from the Kubo formula. It is noted that the expression of the polarization tensor used below is consistent with those derived in the study of QED. \cite{bechler} 

The correlation function for the four-vector current $j^{\mu}(\mathbf{q},\omega)=(c^* \rho(\mathbf{q},\omega), \mathbf{j}(\mathbf{q},\omega))$ is given by
\begin{equation}
\mathopen<j^\mu(\mathbf{q},\omega)j^v(-\mathbf{q},\omega)\mathclose>=-\, \hbar e^2 c^{*\, 2} \Pi^{\mu v}(\mathbf{q},\omega).
\label{correlation}
\end{equation}
The polarization tensor $\Pi^{\mu v}(\mathbf{q},\omega)$ for Dirac electrons is  \cite{Maebashi} 
\begin{equation}
\Pi^{00}(\mathbf{q},\omega+i0^+)=\frac{\epsilon_0}{e^2}q^2\,\chi_e(q,\omega),
\label{pi}
\end{equation}
\begin{equation}
\Pi^{0i}(\mathbf{q},\omega+i0^+)=\Pi^{i0}(\mathbf{q},\omega+i0^+)=\frac{\epsilon_0}{e^2}\frac{\omega}{c^*}q_i\,\chi_e(q,\omega),
\end{equation}
\begin{equation}
\begin{multlined}
\Pi^{ij}(\mathbf{q},\omega+i0^+)= \\
\frac{\epsilon_0}{e^2}\left[\frac{\omega^2}{c^{*2}}\delta_{ij}\,\chi_e(q,\omega)+\frac{c^2}{c^{*2}}(q^2\delta_{ij}-q_i\, q_j)\chi_m(q,\omega)\right].
\label{pip}
\end{multlined}
\end{equation}
As seen in Eqs. (\ref{pi})-(\ref{pip}), the polarization tensor is expressed by two scalar functions $\chi_e(q,\omega)$ and $\chi_m(q,\omega)$, corresponding to the dielectric susceptibility and the magnetic susceptibility, respectively. 
Substituting Eqs.  (\ref{correlation})-(\ref{pip}) into (\ref{Hx}) and (\ref{Hy}), we obtain
\begin{equation}
\begin{multlined}
\mathrm{Im}\left[\mathopen|<\delta\mathbf{H}_x(\mathbf{q},\omega)\delta\mathbf{H}_x(\mathbf{-q},\omega)>\mathclose|+\mathopen|<\delta\mathbf{H}_y(\mathbf{q},\omega)\delta\mathbf{H}_y(\mathbf{-q},\omega)>\mathclose|\right]\\
=-\hbar\epsilon_0\, \frac{q^2+q_z^2}{q^2}\,\mathrm{Im}\left(c^2\chi_m+\frac{\omega^2}{q^2}\chi_e\right).
\end{multlined} 
\end{equation}
The finite temperature expression for Im$\,\chi_m(q,\omega)$ and  Im$\,\chi_e(q,\omega)$ has been obtained as\cite{Maebashi}
\begin{equation}
\begin{multlined}
\mathrm{Im}\, \chi_m(q,\omega)=-sgn(\omega)\frac{c^*}{c}\frac{\alpha}{q^3}  \theta(Q^2-4m^{*\,2}c^{*\,2}/\hbar^2) \\
\times \int_{0}^{\,\lambda}dp_0 \,h(\hbar p_0c^* )\left\{q^2-4p_0^2+\frac{2Q^2}{q^2}(\lambda^2-3p_0^2)\right\} \\
+sgn(\omega)\frac{c^*}{c}\frac{\alpha}{q^3}  \theta(-Q^2) \\
\times\int_{\lambda}^{\infty}dp_0 \,h(\hbar p_0c^* )
\left\{q^2-4p_0^2+\frac{2Q^2}{q^2}(\lambda^2-3p_0^2)\right\} ,
\end{multlined}
\end{equation}
\begin{equation}
\begin{multlined}
\mathrm{Im}\, \chi_e(q,\omega)=sgn(\omega)\frac{c}{c^*}\frac{\alpha}{q^3}  \theta(Q^2-4m^{*\,2}c^{*\,2}/\hbar^2) \\
\times \int_{0}^{\,\lambda}dp_0 \,h(\hbar p_0c^* )\left\{q^2-4p_0^2\right\} \\
 -sgn(\omega)\frac{c}{c^*}\frac{\alpha}{q^3}  \theta(-Q^2)\int_{\lambda}^{\infty}dp_0 \,h(\hbar p_0c^* ) \left\{q^2-4p_0^2\right\},
\end{multlined}
\end{equation}
\begin{equation}
\begin{multlined}
h(x)=\frac{1}{2}\{ f(x-\hbar\left|\omega\right|/2)- f(x+\hbar\left|\omega\right|/2) \\
+f(-x-\hbar\left|\omega\right|/2)- f(-x+\hbar\left|\omega\right|/2)\}\nonumber,
\end{multlined}
\end{equation}
where $Q^2=\omega^2/c^{*\,2}-q^2$, $\lambda=\frac{q}{2}\sqrt{1-\frac{4m^{*\,2}c^{*\,2}}{\hbar^2Q^2}}$ , $f(x)=\left(e^{\beta\,(x-\mu)}+1\right)^{-1}$, $\alpha=\frac{e^2}{4\pi\epsilon_0\hbar c}$ and $\theta(x)$ is the Heaviside step function. The mass term 2$m^{*}c^{*\,2}$ represents the band gap. The term constrained by $\theta(Q^2-4m^{*\,2}c^{*\,2}/\hbar^2)$ corresponds to the electron/hole pair creation, so it can be interpreted as the interband contribution. The term constrained by $\theta(-Q^2)$ corresponds to the gapless electron excitation in the conduction/valence band, hence the intraband contribution.
Finally, we obtain the expression of $1/T_1$ as follows.
\begin{equation}
\begin{multlined}
\frac{1}{T_1}=\lim_{\omega\to0}\frac{\gamma^2_n}{2}\frac{k_B T}{(2\pi)^2}\left(-\frac{8}{3}\hbar\epsilon_0\mu_0^2\, \right)  \int_{0}^{\infty}dq\,\, q^2\frac{\mathrm{Im}\left(c^2\chi_m+\frac{\omega^2}{q^2}\chi_e\right)}{\hbar \omega} \\
=\lim_{\omega\to0}\frac{\gamma^2_n}{2}\frac{k_B T}{(2\pi)^2}\left(-\frac{8}{3}\hbar\epsilon_0\mu_0^2\, cc^*\right)  \int_{0}^{\infty}dq\,\, q^2\frac{\mathrm{Im}\,\chi(q,\omega)}{\hbar \omega}, \label{final}\quad\quad 
\end{multlined}
\end{equation}
where
\begin{equation}
\begin{multlined}
\mathrm{Im}\, \chi(q,\omega)=sign(\omega)\theta(Q^2-4m^{*\,2}c^{*\,2}/\hbar^2)\frac{\alpha　Q^2}{q^5} \\
\times \int_{0}^{\,\lambda}dp_0 \,h(\hbar p_0c^* )\left\{q^2+2(p_0^2-\lambda^2)\right\}\\
 -sign(\omega)\theta(-Q^2) \frac{\alpha　Q^2}{q^5} \int_{\lambda}^{\infty}dp_0 \,h(\hbar p_0c^* )\left\{q^2+2(p_0^2-\lambda^2)\right\}. \label{chi}
\end{multlined}
\end{equation}

In the following section, we evaluate $1/T_1$ using the parameters expected in experiments. Before showing the numerical results, let us consider some limiting cases. 
At low temperatures, we can show that the $T$ linear dependence is obtained due to gapless excitations from the Fermi surface.  In this limit, the interband contribution vanishes regardless of the existence of energy gap. So, we only consider the intraband term at $0$K, assuming the positive chemical potential $\mu$.
In the case of massive Dirac fermions ($m^*c^{*\,2}>>\hbar\omega\approx 0$ and $\lambda\approx\frac{1}{2}\sqrt{q^2+4m^{*\,2}c^{*\,2}/\hbar^2}$), we obtain
\begin{eqnarray}
\frac{1}{T_1 k_B T}&=& 0\quad  \quad(\mu< m^*c^{*\,2})\nonumber \\ 
&=&\frac{1}{2\hbar \pi^3} \left(\frac{\gamma_n e \mu_0}{2} \frac{\sqrt{\mu^2-m^{*2}c^{*4}}}{\hbar c^*}\right)^2\label{low} \\
&\times&\left(1+\frac{2}{3} \mathrm{ln} \left(\frac{2\sqrt{\mu^2-m^{*2}c^{*4}}}{\hbar \omega}\right)\right)\quad(\mu\geq m^*c^{*2}). \nonumber
\end{eqnarray}

We could obtain the expression for massless Dirac fermions by letting $m^*c^{*2}=0$. As $\omega\neq 0$ in experiments, the value of $1/T_1T$ is a non-zero constant, so it has the same temperature dependence as the Korringa relation. In the limit of  $\omega\to 0$, the logarithmic divergence appears for both cases, which is consistent with the result of free electron model.\cite{knigavko_divergence_2007}  We should note that the simple relationship between the Knight shift and $1/T_1T$ does not hold for orbital contributions, if the Knight shift is given by the orbital susceptibility derived for the Dirac Hamiltonian.\cite{fuseya_transport_2015}

At high temperatures, $1/T_1$ shows a different behavior. When the temperature is higher than max($m^*c^{*2},\sqrt{\mu^2-m^{*2}c^{*4}}$), thermal excitations broaden the delta function and the upper cutoff in the integration over $q$ is replaced from $\mu$ to $k_B T$. In this case, we can extract the temperature dependence by changing the integration variable as $\bar{q}=\beta q$,$\bar{p}_0=\beta p_0$ and $\bar{\omega}=\beta \omega$. From Eqs. (\ref{final}) and (\ref{chi}), we obtain
\begin{equation}
\begin{multlined}
\frac{1}{T_1}\varpropto T\int dq\frac{Q^2}{q^3}\int dp_0\, \frac{dh(\hbar p_0c^* )}{d\omega}p_0^2\nonumber\\
= T^3  \int d\bar{q}\frac{\bar{Q}^2}{\bar{q}^3}\int d\bar{p}_0\, \frac{dh(\hbar  \bar{p}_0c^*)}{d\bar{\omega}}\bar{p}_0^2. 
\end{multlined}
\end{equation}
Hence, we should expect the transition from $T$ linear to $T^3$ dependence as the temperature is increased. This is confirmed numerically in the following.

To calculate 1/$T_1$ numerically, we define the parameters comparable to the experiments. The value of $c^*$ is approximated to $3\times10^{5} \mathrm{m s^{-1}}=0.001\times \,\mathrm{c}$. A good accuracy in the estimate of $\omega$ is not required as it only affects the result logarithmically. In the following, we approximate $\hbar \omega=10^{-2}$meV.  As the energy gap for Dirac electron systems is typically in the order of $10$meV, $Q^2-4m^{*\,2}c^{*\,2}/\hbar^2=\omega^2/c^{*\,2}-q^2-4m^{*\,2}c^{*\,2}/\hbar^2<0$ for all values of $q$, and the first term in Eq. (\ref{chi}) can be ignored as long as we consider the massive Dirac systems. For simplicity, the mass term $m^*c^{*2}$ is represented by the energy gap $\Delta$ below.

\begin{figure}[b]
\includegraphics[width=90mm]{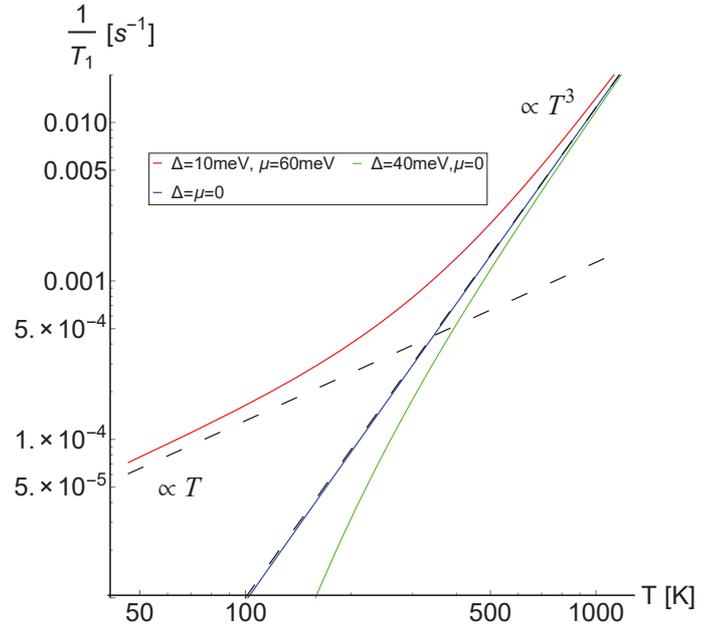}
\caption{(Color online) Plot of $1/T_1$ in Log scale as a function of $T$ at $\hbar \omega=10^{-2}$meV. $\Delta$ represents the energy gap, and $\mu$ is the chemical potential.}
\label{fig:log}
\end{figure}
\begin{figure}[b]
\includegraphics[width=90mm]{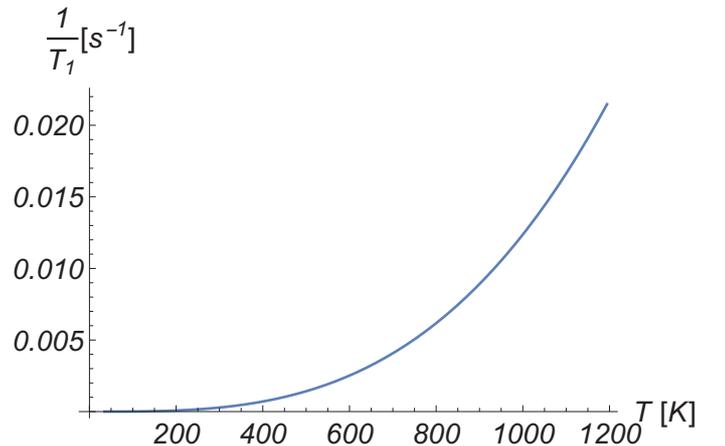}
\caption{(Color online) Plot of $1/T_1$ as a function of $T$ at $\Delta=10$meV,  $\mu=0$.}
\label{fig:relax}
\end{figure}

In Fig.\ref{fig:log}, we plot $1/T_1$ as a function of temperature on a log scale for several values of $\Delta$ and $\mu$. If the chemical potential is above the energy gap, i.e., $\mu>\Delta$,  $1/T_1$ shows a metallic contribution that is proportional to $T$ at low temperatures(red solid line). If the chemical potential is inside the gap, i.e., $\mu<\Delta$, it shows an exponential decrease as known for insulators (green solid line). As we expect, the $T^3$ dependence always appears when the temperature is above max($\Delta,\sqrt{\mu^2-\Delta^2}$). This anomalous temperature dependence occurs in Dirac electron systems because of the small energy gap allowing the thermal excitation from the lower band. For a massless system with the chemical potential on top of the Dirac point, i.e., $\mu=\Delta=0$, $1/T_1$ is proportional to $T^3$ for all temperatures to a good approximation (blue solid line). There is no region where $1/T_1$ is proportional to $T$, because the density of state is zero at the Fermi surface.  

Figure \ref{fig:relax} shows a numerical calculation of $1/T_1$ for $\Delta=10$meV. As discussed above, the expression for $1/T_1$ can be obtained by replacing $\mu$ with $k_B T$ in Eq. (\ref{low}) at high temperatures. Here, it is numerically fitted to a function proportional to $T^3 (1+\frac{2}{3} \mathrm{ln}(4k_B T/\hbar \omega))$. We obtain the expression for the high temperature limit as follows.
\begin{equation}
\frac{1}{T_1k_B T}=1.4\times \frac{1}{2\hbar \pi^3} \left(\gamma_n e\mu_0  \frac{k_B T}{\hbar c^*}\right)^2 \left(1+\frac{2}{3}\mathrm{ln}\left(\frac{4k_B T}{\hbar \omega}\right)\right). 
\end{equation}

\begin{table}
\begin{center}
\begin{tabular}{|l|c|c|} \hline
  & Dimensionality & Temperature dependence of $1/T_1$ \\ \hline
\multirow{2}{*}{Spin(Fermi contact)} & 2D & $T^3$  \\  & 3D & $T^5$ \\ \hline
\multirow{2}{*}{Orbital} & 2D & 0  \\   & 3D & $T^3$ \\ \hline 
\end{tabular}
\end{center}
\caption{Summary of the theoretical results for the temperature dependence of the relaxation rate at the high temperature limit.}
\label{table:summary}
\end{table}

As a summary, we present a comparison between the orbital and the spin contributions to $1/T_1$ in Table \ref{table:summary}. The spin relaxation rate is given by the Fermi contact interaction as discussed above, which can be written as\cite{katayama}
\[\frac{1}{T_1}=\pi T  \int_{-\infty}^{\infty} d\epsilon \left(D(\epsilon)\right)^2 \left(-\frac{\partial f}{\partial \epsilon}\right).\]
Here we should note that the density of states of Dirac electron systems $D(\epsilon)$ depends on dimensions of the system. In $3$D, it is quadratic to the energy. Therefore, $1/T_1$ is proportional to $T^5$, while it is proportional to $T^3$ in $2$D at high temperatures. 
 
Let us compare the present theoretical calculation with experiments. As for the $\beta$-NMR measurement of $\mathrm{Bi}_{0.9}\mathrm{Sb}_{0.1}$\cite{macfarlane_beta-detected_2014}, the $T^3$ dependence at the temperature above the energy gap could explain the rapid increase in $1/T_1$ observed for $\mathrm{Bi}_{0.9}\mathrm{Sb}_{0.1}$ at 225K. However, the experimental data points are not enough to determine whether this increase in $1/T_1$ is proportional to $T^3$ or not. As no sharp increase was found for Bi, we could expect the energy gap of Bi to be greater than the measured temperature range.

\begin{figure}[b]
\includegraphics[width=90mm]{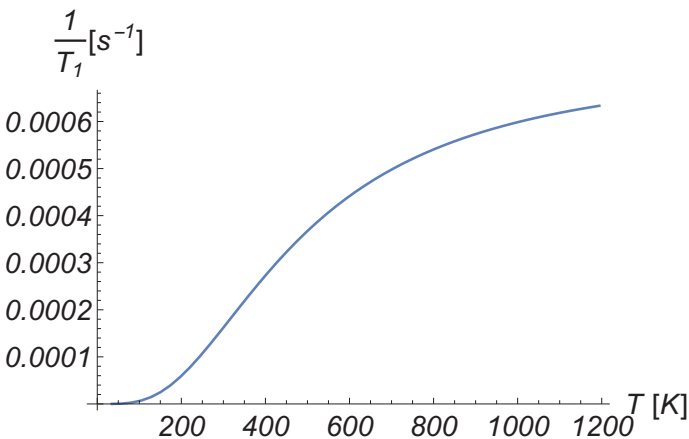}
\caption{(Color online) Plot of $1/T_1$ as a function of $T$ at $\hbar c^*p_0^{max}=100$ meV, $\Delta=10$meV and $\mu=0$.}
\label{fig:cutoff}
\end{figure}

The major difference with our result is that the sharp rising in 1/$T_1$ of $\mathrm{Bi}_{0.9}\mathrm{Sb}_{0.1}$ was found only in a small region of temperatures. This is probably due to the limitation of the Dirac Hamiltonian model, because the approximation to the Dirac Hamiltonian can be verified only near the Dirac point for real materials. To take account of this effect, we introduce an upper bound for the energy of Dirac electrons in the integral over $p_0$ of Eq.(\ref{chi}). In this case, the $T^3$ dependence is suppressed at higher temperatures, as shown for $\hbar c^*p_0^{max}=100$meV in Fig.\ref{fig:cutoff}.  This result is consistent with the temperature dependence of $1/T_1$ observed in the $\beta$-NMR experiment.  To improve the precision of the order of magnitude, we would need to include the effect of anisotropy, but this is beyond the scope of this paper.

For three-dimensional Dirac electron systems as well as Weyl semimetals, we could distinguish the spin/orbital contributions by the temperature dependence of $1/T_1$ at high temperatures. \cite{NQR_TaP,weyl} Possible candidates for measurements are the cubic inverse perovskites such as Sr$_3$PbO, as predicted by the first-principles calculation.\cite{kariyado} The NMR experiment was carried out on Sr$_3$PbO recently. \cite{suetsugu} However, as pointed out by Suetsugu et al.,\cite{suetsugu} the spin contribution can also show the crossover from $T$ linear to $T^3$ dependence around room temperature in the samples with large carrier densities due to the Sommerfeld expansion for $\sqrt{\mu^2-\Delta^2} \gtrapprox k_B T$. In contrast, $1/T_1$ from the orbital effect becomes linear in $T$ when the carrier density is high and $\sqrt{\mu^2-\Delta^2}>k_B T$. To see the crossover from $T$ linear to $T^3$ dependence, it is necessary to measure in the parameter range of $\sqrt{\mu^2-\Delta^2}<k_B T$.

For quasi-two-dimensional systems, the $T^3$ dependence of relaxation time has been understood as the contribution from the Fermi contact interaction of the two-dimensional gapless state of $\alpha$-(BEDT-TTF)$_2 \mathrm{I}_3$\cite{miyagawa}. In addition, the orbital contribution to $1/T_1$ will exist, which can be discussed in the similar approach developed in the present paper. However, it should be noted that the electronic state in $\alpha$-(BEDT-TTF)$_2 \mathrm{I}_3$ is in the quantum limit in the NMR experiments. Thus, it will be necessary to take account of the effect of Landau quantization, which remains as a future problem.

In this paper, we considered the effect of current-induced magnetic field on the nuclear spin relaxation time. Our work suggests that the orbital contribution should not be neglected to understand the peculiar behavior of the nuclear relaxation time of Dirac electron systems, and more generally of the systems with a small effective electronic mass.

TH would like to thank H. Matsuura, T. Mizoguchi, H. Yasuoka, S. Suetsugu, K. Kitagawa and H. Takagi for fruitful discussions and comments. This work was supported by Japan Society for the Promotion of Science through Program for Leading Graduate Schools (MERIT) and a Grant-in-Aid for Scientific Research on ``Multiferroics in Dirac electron materials''(No.15H02108).

\end{document}